\documentclass[superscriptaddress,twocolumn,pre]{revtex4}
\usepackage{amsmath}
\usepackage{amsfonts}
\usepackage{amssymb}
\usepackage{graphicx}
\graphicspath{{images/}} 
\usepackage{color}
\usepackage[pdfstartview=FitH,
            breaklinks=true,
            bookmarksopen=false,
            bookmarksnumbered=true,
            colorlinks=true,
            linkcolor=black,
            citecolor=black,
            urlcolor=black,
            pdftitle={Evolution of immune systems},
            pdfauthor={AM, TM, OR, and AW},
            pdfsubject={Paper}
            ]{hyperref}

\newcommand{\B}{\boldsymbol}

\begin{document}

\title{Diversity of immune strategies explained by adaptation to
pathogen statistics}

\author{Andreas Mayer} 
\affiliation{Laboratoire de physique th\'eorique,
    CNRS, UPMC and \'Ecole normale sup\'erieure,
    75005 Paris, France}
\author{Thierry Mora}
\affiliation{Laboratoire de physique statistique,
    CNRS, UPMC and \'Ecole normale sup\'erieure,
    75005 Paris, France}
\author{Olivier Rivoire} 
\affiliation{Laboratoire Interdisciplinaire de Physique,
    CNRS and Universit\'e Grenoble Alpes,
    38000 Grenoble, France}
\author{Aleksandra M. Walczak}
\affiliation{Laboratoire de physique th\'eorique,
    CNRS, UPMC and \'Ecole normale sup\'erieure,
    75005 Paris, France}

\address{}

\date{\today}

\begin{abstract}
     % abstract
Biological organisms have evolved a wide range of immune mechanisms to defend themselves against pathogens. Beyond molecular details, these mechanisms differ in how protection is acquired, processed and passed on to subsequent generations -- differences that may be essential to long-term survival. Here, we introduce a mathematical framework to compare the long-term adaptation of populations as a function of the pathogen dynamics that they experience and of the immune strategy that they adopt. We find that the two key determinants of an optimal immune strategy are the frequency and the characteristic timescale of the pathogens. Depending on these two parameters, our framework identifies distinct modes of immunity, including adaptive, innate, bet-hedging and CRISPR-like immunities, which recapitulate the diversity of natural immune systems.

\end{abstract}

\maketitle

 % intro

Immune systems have evolved to protect organisms against large and unpredictable pathogenic environments. Yet immunity always comes at a cost (metabolic and maintenance costs, auto-immune disorders, etc.~\cite{schmid-hempel-2005}), and this cost must be balanced by the benefits that protection confers~\cite{deerenberg-2000, clark-2008}. Faced with the problem of evolving a suitable defense, different organisms, from archae to humans, have developed different strategies to identify and target pathogens, which have given rise to a diversity of mechanisms of immunity.

 A lot of effort has been put into elucidating these mechanisms down to their molecular details in a variety of species~\cite{janeway-1997, magnadottir-2006, dangl-2006, strand-2008, moineau-2010, sontheimer-2010}.
 Beyond many differences, these studies have revealed many commonalities \cite{boehm-2011, read-2005}, which hint at a possible general understanding of the trade-offs that shape their design \cite{deerenberg-2000, schmid-hempel-2005}. 
For instance, independently of the well-known adaptive immune systems of jawed vertebrates, jawless vertebrates (e.g. lampreys) have developed an alternative adaptive system that uses a distinct molecular family of receptors, but both systems function largely in the same way, relying on the generation of a large number of diverse receptors expressed by two types of lymphocytes (B or T-like cells). Likewise, the innate immune systems of invertebrates and vertebrates, share many similarities, relying on the selected expression of germline Toll-like receptors upon infection. Some of the features of vertebrate immunity are even shared  with bacteria, who have developed their own targeted immunity based on the CRISPR/Cas system \cite{barrangou-2010,sontheimer-2010}, which itself bears strong resemblance with genome protection through interfering RNAs in eukaryotes \cite{hannon-2009}.

Independently of how they evolved and their particular molecular implementation, we may classify these diverse mechanisms into a few broad modes of immunity: 
heritable but not adaptable within an individual’s lifetime, as innate immune systems; heritable and adaptable within a lifetime but with the benefits of adaptation being non heritable, as adaptive immune systems; acquired from the environment and heritable, as the CRISPR/Cas system; and mixed strategies combining several of these elements.
These broad distinctions call for general principles to characterize the conditions under which one or another mode of immunity may be expected to evolve \cite{boehm-2011, read-2005, schmid-hempel-2005}. 
The diversity and variability of threats from the pathogenic environment suggests that different modes of immunity may offer better protection depending on
the patterns of occurrence of pathogens or the effective population size of the protected population. Here we apply a general theoretical framework for analyzing populations in a varying environment \cite{leibler-2014} to predict the emergence of the basic forms of observed immunity.

 % figure1
\begin{figure*}
    \centering
    \includegraphics[width=\linewidth]{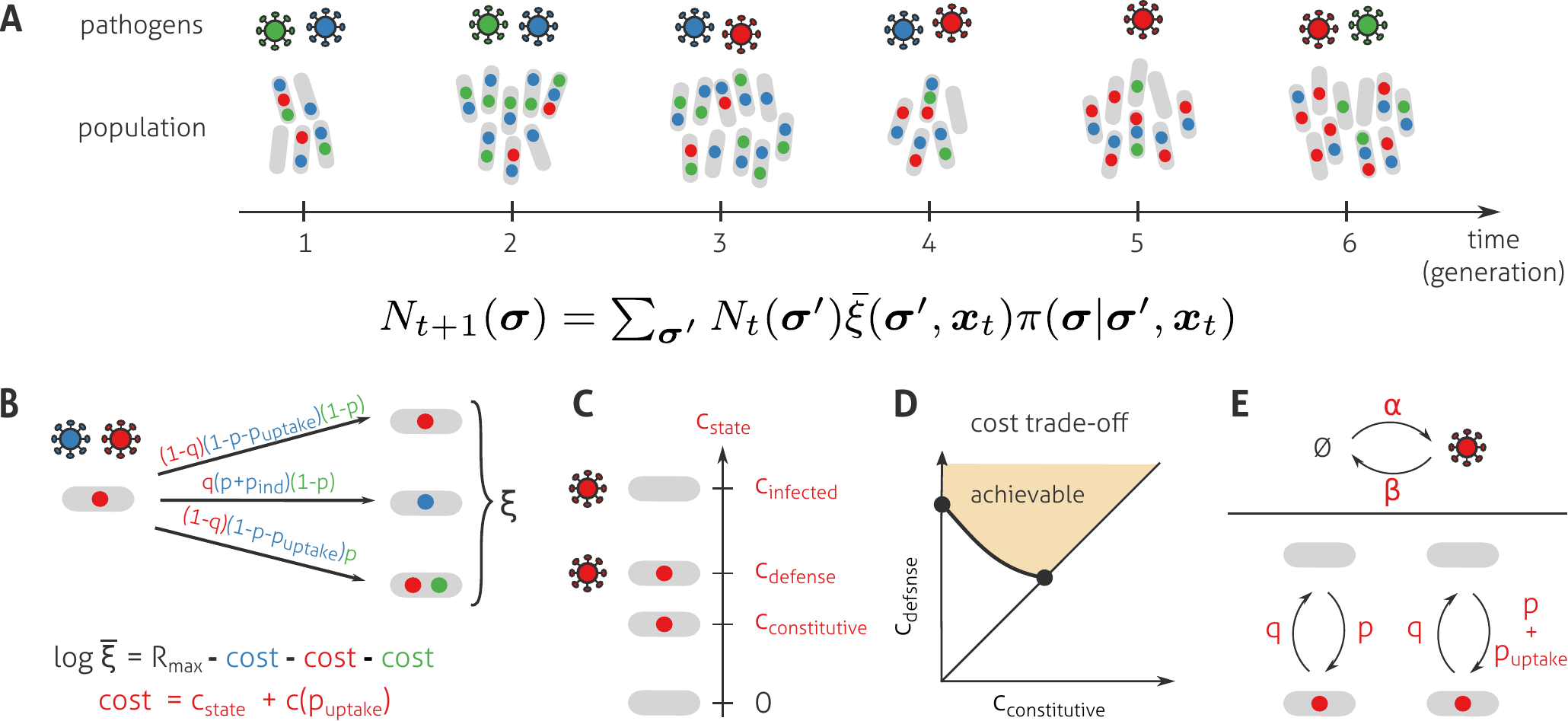}
    \caption{
A model to explore the incidence of different modes of immunity on the long-term growth of populations.
        {\bf (A)}~A population of organisms, each possibly protected against no, one or several pathogens (no, one or several colored dots) evolves in presence of a pathogenic environment that varies from generation to generation. The mean number of individuals with protection $\sigma$ at generation $t$, $N_{t} (\sigma)$, is given by a recursion equation involving the mean number of offspring $\bar\xi(\sigma', x_t)$ for individuals with protection $\sigma'$ and the probability $\pi(\sigma|\sigma', x_t)$ that each of their offspring inherits a protection $\sigma$; both of these quantities may depend on the current pathogenic environment $x_t$. The long-term growth rate of the population is given by $(1/t)\ln N_t$ at large $t$, with $N_t=\sum_\sigma N_{t}(\sigma)$ the total population size.
          {\bf (B)} The $\xi$ offspring produced by an individual inherits the immune protections of their parent with rules specified in panel E. Each pathogen reduces the mean number of offspring $\bar\xi(\sigma, x_t)$ by a cost $c_{\rm state}$ that depends on whether the individual is in state `infected', `defense', or `constitutive' relative to the pathogen, and by a cost $c(p_{\rm uptake})$ that depends on the rate $p_{\rm uptake}$ at which protection is directly induced by the presence of the pathogen.
      {\bf (C)}  An unprotected organism pays a cost of infection $c_{\rm infected}$ if the pathogen is encountered, which is reduced to $c_{\rm defense}$ if it is protected. A protected organism must, however, pay a constitutive cost $c_{\rm constitutive}$ even in the absence of the pathogen, while an unprotected organism pay no cost.
      {\bf (D)} We assume a trade-off between the constitutive and defense costs: a more efficient defense (lower $c_{\rm defense}$) requires more resources (higher $c_{\rm constitutive}$). 
      {\bf (E)} Dynamics of appearance and disappearance of pathogens $x_t$ and immune protection $\sigma$.
A pathogen appears with rate $\alpha$ and disappears with rate $\beta$; these rates define the frequency $\pi_{\rm env}=\alpha/(\alpha+\beta)$ and characteristic time $\tau_{\rm env} = -1/\ln(1 - \alpha - \beta)$ of the pathogen. Protection against a given pathogen is acquired spontaneously with rate $p$, and lost from one generation to the next with rate $q$. Additionally, the presence of the pathogen can increase the rate of acquisition of protection by $p_{\rm uptake}$, as {\em e.g.} in the CRISPR/Cas system of prokaryotes.  
    \label{fig1}}
\end{figure*}

 % setup

Individuals reproduce in the presence of pathogens, which randomly appear, may persist for several generations and disappear before possibly reappearing at a latter time (Fig.~1A). In our framework, a given pathogen has a probability $\alpha$ to appear and a probability $\beta$ to disappear from one generation to the next (Fig.~1E). The pathogenic dynamics is quantified both by the pathogen frequency $\pi_{\rm env} = \alpha/(\alpha+\beta)$, which is the probability that it is present at any given generation, and by the characteristic timescale $\tau_{\rm env}= -1/\ln(1 - \alpha - \beta)$, which sets how fast pathogens appear and disappear.

Pathogens reduce the fitness of the individuals in the population and the immune system is designed to mitigate this effect. An individual's fecundity, defined as its expected number of descendants in the next generation $\bar\xi$, depends on the pathogenic environment and its ability to protect itself against it. Each pathogen independently lowers the fecundity of an unprotected individual by a relatively large cost factor $c_{\rm infected}>0$ (Fig.~1B). This cost is reduced to a lower cost $c_{\rm defense}<c_{\rm infected}$ when the individual is protected by its immune system, however this protection comes at a minimal but ever-present cost of $c_{\rm constitutive}<c_{\rm defense}$ of maintaining the immune defense in absence of the pathogen (Fig.~1C). 

We explore the choices and tradeoffs underlying various modes of immunity along three axes: adaptability, heritability, and mode of aquisition. The first, adaptability axis concerns how much resources are invested in the protection for the return of an efficient response. This tradeoff imposes a relationship between $c_{\rm defense}$ and $c_{\rm constitutive}$ (Fig.~1D):
the more effective the defense (the lower $c_{\rm defense}$), the higher  maintenance cost (the higher $c_{\rm constitutive}$). For example, having a large number of immune cells specialized against a specific pathogen allows for a quick and efficient response in case of invasion, but this enhanced protection comes at the cost of producing and maintaining these cells in the absence of the pathogen. This strategy is adopted, for example, by much of the innate immune systems of plants and animals \cite{janeway-1997}. On the contrary, the adaptive immune system keeps a very small specialized pool of lymphocytes for each potential antigen, and makes them proliferate only in case of infection \cite{boehm-2011}. The second, heritability axis is defined by the probability $q$ that the protection is not transmitted to the offspring (Fig.~1E). Finally, the third, acquisition axis specifies how individuals may acquire the protection without inheriting it from their parent. This acquisition may occur randomly independently of the environment, with probability $p$, for instance by mutation or phenotypic switching, as is the case for antibiotic resistance in bacteria \cite{gniadkowski-2008}; or it can be induced by the presence of the pathogen with probability $p_{\rm uptake}$, as in CRISPR-Cas immunity (Fig.~1E) \cite{sontheimer-2010}. This mechanism comes at an extra cost $c(p_{\rm uptake})$ due to maintenance and the risks of uptaking foreign genetic material (Fig.~1B).

 % figure2
\begin{figure*}
    \centering
    \includegraphics[width=\linewidth]{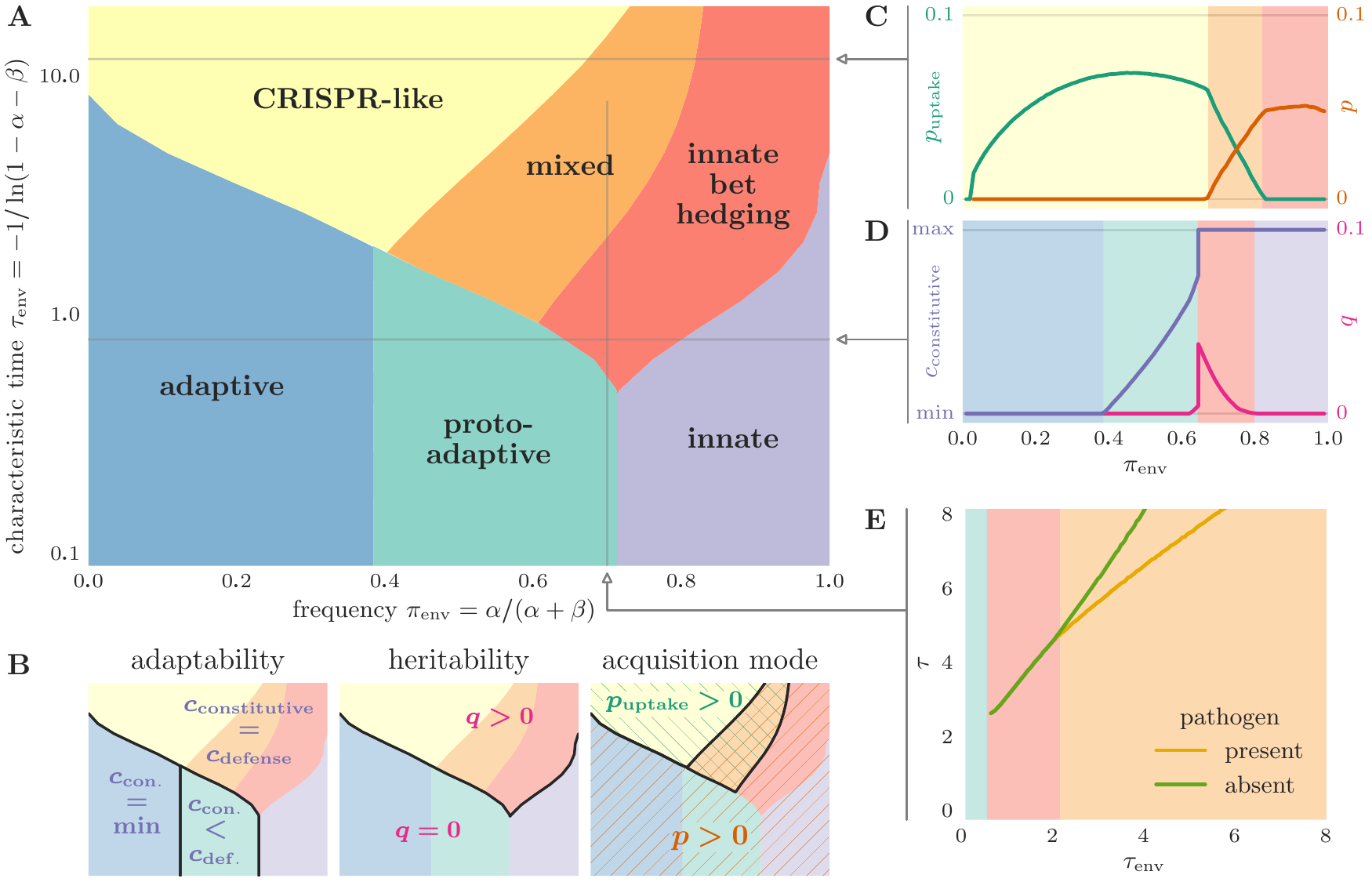}
    \caption{Optimal immune strategies as a function of the  frequency and characteristic time of pathogens.
        {\bf (A)} Distinct optimal immune strategies emerge for different statistics of appearance of the pathogens. 
        Each phase is characterized by the value of parameters indicated in panel B and named after a known immune system that has similar characteristics.
        {\bf (B)} The different phases of immunity are defined by the values of parameters along three main axes: adaptability (constitutive cost $c_{\rm constitutive}$), heritability ($1-q$) and mode of acquisition ($p$ and $p_{\rm uptake}$). 
        {\bf (C)} and {\bf (D)} Optimal parameters as a function of $\pi_{\rm env}$ for $\tau_{\rm env} = 12$ (C) and  $\tau_{\rm env} = 0.8$ (D). For slowly varying environments (C), rare pathogens are best targeted by CRISPR-like uptake of protection, while frequent pathogens are best dealt with by spontaneous acquisition of protection, with a crossover in-between where both co-exist.
	For faster varying environments (D), the constitutive cost invested in the protection goes from negligible to maximal as the pathogen frequency increases. When it is maximal, the best strategy transitions from bet-hedging ($q>0$) to a full protection of the population ($q=0$).
	{\bf (E)} The correlation times of protection in absence of the pathogen, $\tau = -1/\ln(1 - p - q)$, and in its presence, $\tau = -1/\ln(1 - p - p_{\rm uptake} - q)$, are shown for $\pi_{\rm env} = 0.7$ as a function of $\tau_{\rm env}$. Both increase with the correlation time of the pathogen. 
        In this figure, an infinite population size is assumed and the following choices are made: $c_{\rm infection} = 3; \; c_{\rm constitutive} = \left(1.8 - c_{\rm defense}\right) / \left(c_{\rm defense} - 0.2\right); \; c(p_{\rm uptake}) = 0.1\times p_{\rm uptake} + p_{\rm uptake}^2$ (see Fig.~S2 for other choices).
    \label{fig2}}
\end{figure*}

 % discussion

Each choice of the parameters $c_{\rm constitutive}$, $q$, $p$ and $p_{\rm uptake}$ defines a specific immune strategy. This strategy is optimal if a population that adopts it outgrows in the long run any other population following a different strategy. Our goal is to characterize this optimal strategy, in particular its dependency on the two key properties of the pathogen, its frequency $\pi_{\rm env}$ and its characteristic time $\tau_{\rm env}$.  We achieve this goal by 
maximizing the long-term growth rate of populations, defined by $(1/t) \ln N(t)$, where $N(t)$ is the total population size at generation $t$ (Fig.~1A; see SI for a derivation)~\cite{leibler-2011}. 
Conveniently, since the fecundity is affected independently by the different pathogens, each pathogen contributes additivitely to the growth rate and can 
be studied separately (Fig. 1B and SI).
Remarkably, we obtain qualitatively different optimal solutions for given values of $\pi_{\rm env}$, $\tau_{\rm env}$, 
with sharp transitions between these strategies as one varies the parameters of the pathogen statistics, allowing us to define distinct immune regimes (Fig. 2A). The emergence of these very distinct regimes is not an assumption, but the result of the optimisation itself.

Fig.~2B describes these optimal strategies along the three axes of variation outlined earlier.  Along the first axis of variation, adaptability, we find that frequent or persistent pathogens are best dealt with by constitutively expressed immunity ($c_{\rm constitutive}=c_{\rm defense}$), and rare and transient pathogens by investing minimally in the defense ($c_{\rm constitutive} = 0$, in blue); between these two extremes, only a limited form of adaptation is required ($c_{\rm constitutive}<c_{\rm defense}$, in green). Along the second axis, heritability, we find that carrying the protection at all times ($q=0$) is beneficial for fast pathogens but that losing the protection with probability $q>0$ is more advantageous for slow ones. Finally, along the third axis, acquisition, we verify that there is no need to pay the price of informed acquisition ($p_{\rm uptake}=0$) whenever protection is systematically inherited ($q=0$); when it is not the case ($q>0$), we find that uptake is advantageous for sufficiently infrequent pathogens (yellow and orange regions) but that only for very infrequent pathogens does it becomes the exclusive mode of acquisition of protection ($p=0$, $p_{\rm uptake}>0$, in yellow).

Each of these distinct regimes, or phases, is instantiated by natural immune systems. For transient and rare pathogens (blue phase), the optimal strategy is to inherit a defense with minimal constitutive cost. This strategy is characteristic of the adaptive immune system in vertebrates, where an effective immune response is mounted from a small number of precursor cells, the marginal cost of which is negligible \cite{boehm-2011}. For transient but frequent pathogens (purple phase), the optimal strategy consists instead in inheriting a maximally efficient protection that makes the individuals effectively insensitive to the presence of the pathogen at the expense, however, of a large constitutive cost.
The recognition of pathogen-associated molecular patterns by pattern recognition receptors, as for instance the recognition of lipopolysaccharide by Toll-like receptors, is an example of such an innate strategy \cite{janeway-1997}.
An intermediate phase (in green) separates these two extremes, where adaptation is present with a non-zero constitutive cost.
This strategy, which we call proto-adaptive, is represented by certain specialized cells of the innate immune system,
such as natural killer cells \cite{ugolini-2011},
whose abundance can vary as a function of experienced infections, effectively implementing an adaptive memory within a single generation.

For slow and frequent pathogens (red phase), protection is acquired with probability $p>0$ and lost with probability $q>0$ independently of the presence of the pathogen. This bet-hedging strategy is implemented in bacteria that can switch on or off the expression of phage receptors \cite{moineau-2010}. For slow but unfrequent pathogens (yellow phase), a form of bet-hedging is again present, but this time with a non-zero probability to acquire protection only in presence of the pathogen. An example of such a Lamarckian strategy is the CRISPR-Cas immune system in bacteria \cite{sontheimer-2010}. Finally, a mixed phase (in orange) is also possible where protection is randomly acquired at a rate that is increased by the presence of the pathogen.

It is instructive to examine how the parameters of immunity vary within the phases (Fig.~2C-D and S1). As one may expect, the statistical properties of the protection tend to track the pathogen statistics \cite{walczak-2015}. The more frequent the pathogen, the more prevalent the protection in the population (Fig.~2C). Likewise, the characteristic time of the protection, $\tau$, grows with that of the pathogen, $\tau_{\rm env}$ (Fig.~2E).

The phase portrait of Fig.~2A  rationalizes the salient differences between the immune systems of prokaryotes and vertebrates. Bacterial and archeal pathogens evolve on timescales that are much closer to those of their pathogens than vertebrates. From the viewpoint of microbes, the pathogenic environment is relatively constant ($\tau_{\rm env}> 1$), while for vertebrates a particular pathogenic strain is unlikely to survive a single generation ($\tau_{\rm env}\ll 1$). Consistently with our results, vertebrates use fully heritable modes of immunity, and do not rely on bet-hedging. To deal with unfrequent and fast evolving pathogens such as viruses, they recourse to adaptive mechanisms by which they can upregulate their protection in case of an invasion. The three predicted strategies -- adaptive, proto-adaptive, and innate -- correspond to the known modes of immunity in vertebrates \cite{janeway}. Prokaryotes, on the other hand, almost systematically use bet-hedging strategies. They recourse to both the CRISPR-Cas system of acquired immunity \cite{sontheimer-2010}, and to innate immunity through {\em e.g.} restriction endonucleases \cite{moineau-2010}, which correspond to the predicted Lamarckian and innate bet-hedging strategies of the diagram, respectively.
These results are robust to changes of parameters, although increasing costs can make bet-hedging beneficial even at short characteristic times (SI Fig.~S2).

Bacteria and vertebrates also have very different population sizes, which influence their overall survival probability. To evaluate this impact, we ran stochastic simulations competing different strategies for increasing population sizes (see SI Text and SI Fig.~S3). The phase diagram of Fig.~2A was recovered for populations as small as a thousand, while for smaller populations the boundaries between regimes were smeared. Adaptive strategies were generally favored over CRISPR-like strategies in small populations. In addition, for finite populations it is always beneficial to recourse some degree of bet-hedging in order to react quickly to environmental changes and avoid extinction.

By analyzing the long-term fate of populations under minimal assumptions concerning the rules governing adaptability, heritability and acquisition of immune protections, we have recovered the basic known modes of immunity. Remarkably our results hold even for a  single pathogen. 
The key determinants of optimal immune strategies are found to be the statistical features of pathogen occurrence: its frequency and its characteristic timescale. As an implication, a diverse pathogenic environment, with varying statistics, will favor mixed solutions, consistently with the observation of multiple immune systems within a same organism -- such as adaptive and innate immune systems in vertebrates, or CRISPR and innate defense in bacteria. Naturally, the molecular implementation of these general principles differs greatly even between organisms sharing the same type of immunity.
Yet an evolutionary perspective that accounts for the costs and benefits of protection is enough to explain the most salient features of immunity. It will be interesting to extend our framework to account for other essential features of immunity, {\em e.g.} the acquisition of protection by horizontal transfer or the coevolutionary dynamics between pathogens and their hosts. In view of our analysis, it is already less surprising that complex forms of immunity such as the adaptive immune system have evolved separately in jawed and non-jawed vertebrates, with the same general features but different molecular encodings.

{\bf Acknowledgements.}  % acknowledgement
The work was supported by grant ERCStG n. 306312.

\onecolumngrid
\appendix

 % appendix

\section{Setup of the model and recursion equation}
We mathematically study the dynamics of a population of organisms defending against a stochastic pathogen environment using a combination of various immune strategies.

In the model the pathogenic environment is desribed by a $L$-dimensional vector $\B x$  (here bold symbols refer to vectors) of $0$s and $1$s, where $x_i=1$ if pathogen $i$ is present, $0$ otherwise.
Protection of an organism against these pathogens is also described by a $L$-dimensional vector $\B \sigma$, where $\sigma_i=1$ if the protection (antibody, TCR, CRISPR spacer) against pathogen $i$ is present, and $0$ otherwise.

We consider the dynamics of a population of organisms reproducing at discrete times $t$.
Let $N_t(\boldsymbol \sigma)$ be the mean number of organisms in the population at time $t$ with protection $\B \sigma$, for a given environment history \cite{leibler-2011}.
The change in population composition from one generation to the next is governed by the reproductive success of individuals in each state $\B s$, modulated by state switching from parents to offspring:
\begin{equation} \label{eqfullrecursion}
    N_{t+1}(\boldsymbol \sigma) = \sum_{\boldsymbol \sigma'} N_t(\boldsymbol \sigma') \bar \xi(\boldsymbol \sigma', \boldsymbol x_t) \pi(\boldsymbol \sigma | \boldsymbol \sigma', \boldsymbol x_t),
\end{equation}
where $\bar \xi(\boldsymbol \sigma, \boldsymbol x_t)$ is the mean number of offspring of an organism of type $\B \sigma$ in environment $x_t$
and $\pi(\boldsymbol \sigma | \boldsymbol \sigma', \boldsymbol x_t)$ is the switching probability from protection state $\B \sigma'$ to state $\B \sigma$.
Note that the protection state switching probability, which represents to what extent protection is inherited, acquired or lost, generally depends on the state $\B x_t$ of the environment.

A similar recursion to Eq.~\ref{eqfullrecursion} can be rewritten for the fraction of the population in each state, $n_t(\B s)=N_t(\B \sigma)/N_t$, with $N_t=\sum_{\B \sigma} N_t(\B \sigma)$ the total population size:
\begin{equation} \label{recursion}
n_{t+1}(\boldsymbol \sigma) = \frac{1}{Z_t}\sum_{\boldsymbol \sigma'} n_t(\boldsymbol \sigma') \bar \xi(\boldsymbol \sigma', \boldsymbol x_t) \pi(\boldsymbol \sigma | \boldsymbol \sigma', \boldsymbol x_t),
\end{equation}
where $Z_t$ is a normalization constant enforcing $\sum_{\B \sigma} n_t(\B \sigma)=1$.
The population size verifies $N_t=N_0\prod_{t'=0}^{t-1} Z_{t'}$, so that 
the long-term growth rate, $\Lambda = \lim_{T \to \infty}\frac{1}{T} N_T$, is given by:
\begin{equation} \label{eqLambda}
\Lambda = \lim_{T \to \infty} \frac{1}{T}\sum_{t = 0}^T \log(Z_t).
\end{equation}
This rate provides a measure of long-term fitness \cite{leibler-2011}.

We assume that the mutation and inheritance probabilities of different pathogen-protection pairs are independent of each other, {\em i.e.} that  $\pi(\B \sigma|\B \sigma',\B x_t)$ factorizes over the pathogens,
\begin{equation}\label{factorizedswitches}
\pi(\B \sigma|\B \sigma',\B x_t)=\prod_i \pi_i(\sigma_i|\sigma_i',x_{i;t}).
\end{equation}
The entries of $\pi_i(\sigma_i|\sigma_i',x_{i;t})$ are given by Fig.~1E of the main text: $\pi_i(1|0,x)=p+xp_{\rm uptake}$, $\pi_i(0|1,x)=q$.

In addition, the effects of different pathogen-protection pairs on the growth rate are taken to be additive, so that, following the definitions of Fig.~1C:
\begin{equation}\label{cost}
\log \bar \xi = R_{\rm max}-\sum_{i=1}^L \left[c_{\rm infection,i}(1-\sigma_i)x_i + c_{\rm constitutive, i} \sigma_i (1-x_i) + c_{\rm defense, i} \sigma_i x_i  + c(p_{\rm uptake, i})\right],
\end{equation}
where $R_{max}$ is the growth rate in absence of any immune cost.
With these assumptions, the distribution $n_t(\B \sigma)$ also factorizes over $i$:
\begin{equation}
n_t(\B \sigma)=\prod_{i=1}^L (r^t_i \sigma_i + (1-r^t_i)(1-\sigma_i)),
\end{equation}
where $r_i^t$ is the fraction of the population having protection $i$ at time $t$.
Plugging this Ansatz into Eq.~\ref{recursion} with Eqs.~\ref{factorizedswitches} and \ref{cost} yields the following recursion for $r_i^t$:
\begin{equation}\label{eqrecr}
    r_i^{t+1}=\frac{(1-r_i^t)e^{-c_{\rm infection,i} x_i^t} \left(p_i + p_{\rm uptake, i} x_i^t\right) + r_i^t e^{-c_{\rm defense, i} x_i^t - c_{\rm constitutive, i}(1-x_i^t)}(1-q) }{\left(1-r_i^t\right)e^{-c_{\rm infection,i} x_i^t}+r_i^t e^{-c_{\rm defense, i} x_i^t - c_{\rm constitutive, i}(1-x_i^t)}}.
\end{equation}
The recursion depends on the sequence of $x_i^t$, which is a stochastic binary process switching from 0 to 1 with probability $\alpha$, and from 1 to 0 with probability $\beta$, as in Fig.~1E of the main text. Note that the sequence $x_i^t$ is the same for the whole population (a quenched variable in the statistical mechanics sense).
We have $Z_t=e^{R_{\rm max}} \prod_{i=1}^L z^t_i$, with:
\begin{equation}
    z^t_i=e^{-c(p_{\rm uptake, i})} \left[\left(1-r_i^t\right)e^{-c_{\rm infection,i} x_i^t}+r_i^t e^{-c_{\rm defense, i} x_i^t - c_{\rm constitutive, i}(1-x_i^t)}\right]
\end{equation}
From Eq.~\ref{eqLambda} it then follows that
\begin{equation}
    \Lambda = R_{\max} + \sum_{i=1}^L \left(\lim_{T\to\infty}\frac{1}{T}\sum_{t=1}^T\log z^t_i\right).
\end{equation}
The long-term growth rate is a sum of independent terms for each pathogen-protection pair, which allows us to treat the problem
of maximizing long-term growth rate one pathogen at a time.

\section{Numerics}
The cost function of the optimization, $\Lambda$, can be approximated by solving the recursion equation (Eq.~\ref{eqrecr}) for a large enough number of generations (we used at least $10^6$ generations). Since the process is ergodic, averaging over very long periods is equivalent to repeating the process multiple times.
The longer the simulation, the more accurate the evaluation of $\Lambda$. Our goal is to optimize $\Lambda$ over the four parameters $p, q, p_{ind}, c_{\rm constitutive}$ constrained to their domain of definition.
For numerical purposes, all four parameters are first mapped onto the unit interval $(0,1)$. The noise in the evaluation of $\Lambda$ makes the optimization challenging.
It can be reduced by prolonged simulation or repeated sampling at the expense of a higher computational cost per function evaluation.
To optimize under these constraints we use a two-phase algorithm.
In the first phase the DIRECT algorithm \cite{stuckman-1993} provides us with a rough, but global optimization for which we use a relatively low quality approximation.
The results of this first phase are then refined by a pattern search algorithm \cite{torczon-2003}.
To minimize the effects of stochasticity, the algorithm takes averages over an adaptive number of independent runs in this step.
By estimating the standard error of the computed long-term growth rate from the independent runs, this number can be adapted so that differences between compared parameter values are statistically significant.
In the same manner we can ascertain that the final set of parameter values is locally optimal within some tolerance.

To obtain a phase diagram such as the one shown in Fig.~2A we performed a global optimization over all four parameter values as described above, for every environment condition $(\pi_{\rm env},\tau_{\rm env})$.
Based on the results of this first step we defined the features of the obtained phases. All phases are defined by a subset of the variables lying at a constraint boundary.
In order to get more precise boundaries with reasonable computational effort we look at the difference in growth rates between pairs of strategies across environmental conditions.
To decrease noise the difference is calculated across pairs of simulations using the same sequence of pathogens $\{x_t\}$. 
Using interpolation, this allows us to obtain the line of transition between the two strategies very precisely.
To prevent e.g. the mixed strategy to reduce to a CRISPR-like strategy we impose that the parameters that are not constrained in a particular strategy are not closer than a tolerance $0.005$ ($0.0005$ for $q$) of the boundary.

\section{Influence of finite population size}
\label{secfinitesize}

To study the influence of the effects of finite population size we perform direct agent-based simulations of a population of adapting individuals with strategies evolving on a slow timescale.
The population has a finite size $N$ that remains fixed over the course of the simulation.
At every generation the parents of the $N$ individuals are drawn from the individuals making up the previous generation with probabilities proportional to the mean number of offspring $\bar \xi$ of these individuals.
The offspring's state $\boldsymbol \sigma$ is determined from the state of its parent $\boldsymbol \sigma'$ according to the switching rates $\pi(\boldsymbol \sigma | \boldsymbol \sigma', x_t)$ defined previously.
Along with the state $\B \sigma$, the switching rates themselves, $\pi(\B \sigma,\B \sigma',\B x)$, as well as the degree of adaptability, $c_{\rm constitutive}$ -- in other words, the parameters defining the immune strategy -- are also transmitted to the offspring. They also change from parent to offspring, although at a much slower rate than the state to preserve a clear separation of timescales between short-term and long-term adaptations.
In this setup, selection acts on the strategies.
After an equilibration phase, we collect statistics on the strategies adopted by individuals in the population.
To get rid of the effect of deleterious mutations that do not eventually fix in the populations the mutation rate and size were scaled down exponentially with time. 
As population size is finite deleterious mutations can fix in the population, which means that even in the limit of zero mutation rate there remains a spread in the distribution of strategies.
Hence we do not only represent the median as a measure of the central tendency of a parameter, but also the interquartile range as a measure of its spread. 
Results are shown in Fig.~\ref{figSIevol}.
For a population size of 1000 the median of strategies follows very closely the optimal strategy for an infinite population.
For smaller population sizes the median starts to deviate from the optimum in the infinite population limit.
One of the most notable changes is an increased adaptability of the strategies as seen in the upward shift of the $c_{\rm constitutive}$ curves with smaller population size at $\pi_{\rm env} = 0.3$.
To lower the chances of all of the population being mal-adapted, the evolved strategies in small populations diversify the protection state more strongly, as seen from the higher mutation rates for smaller population sizes. 
Finally, the quite large interquartile ranges show that as expected significant variation of the evolved strategy coexists in small populations. 

\begin{figure*}
    \centering
    \includegraphics[width=\linewidth]{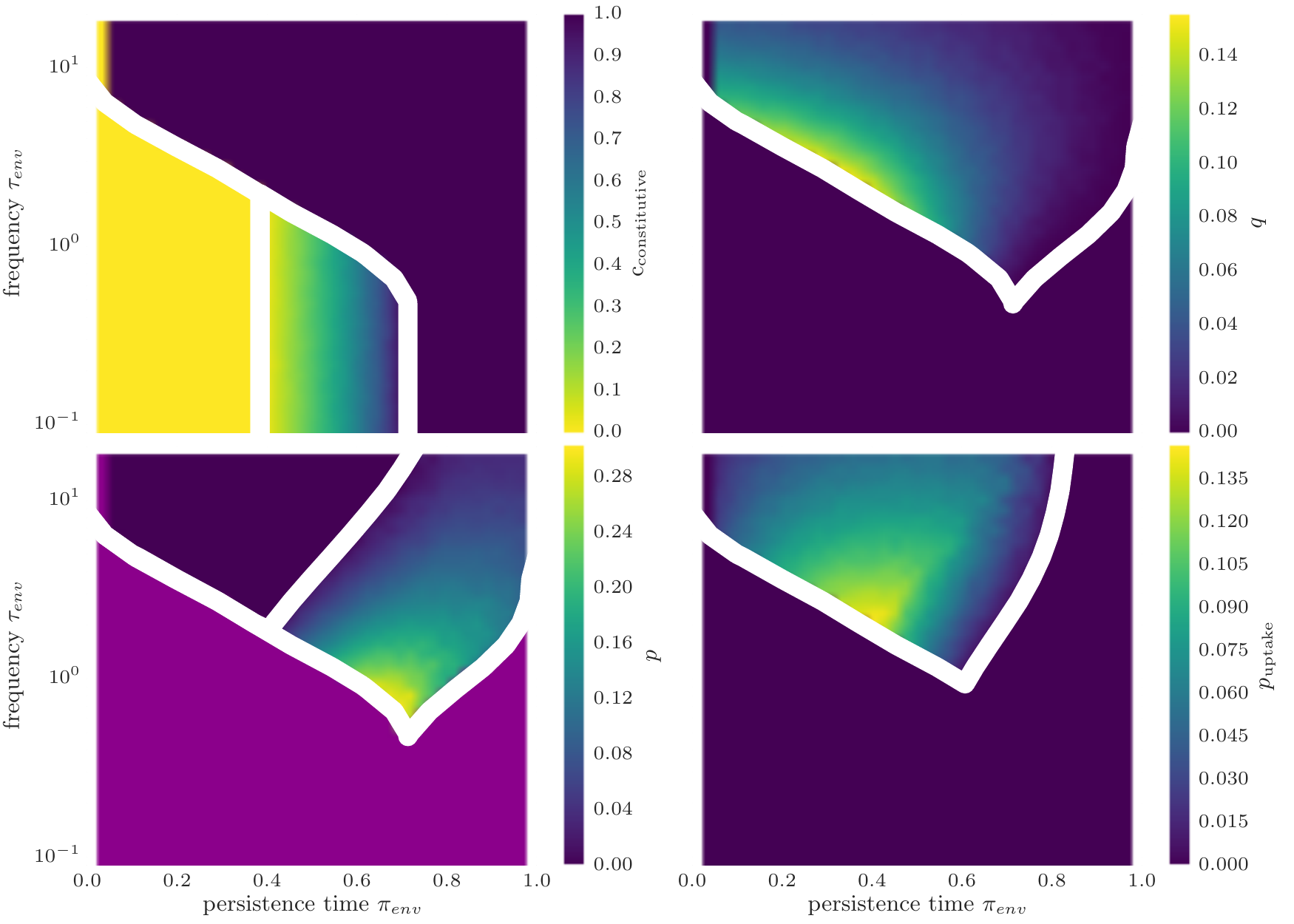}
    \caption{
        Optimal parameters from a global optimization of long-term growth rate. Regions where a parameter is unconstrained at the optimum are shown in purple.
        Phase boundaries pertaining to the shown parameter in white.
        A maximum number of 5000 function evaluations is used for the first phase of the optimization.
        The second phase of the optimization is terminated at a tolerance in the parameter values of $0.005$.
        The same model parameters as in Fig.~2 are used.
    \label{figSIopt}}
\end{figure*}

\begin{figure*}
    \centering
    \includegraphics[width=\linewidth]{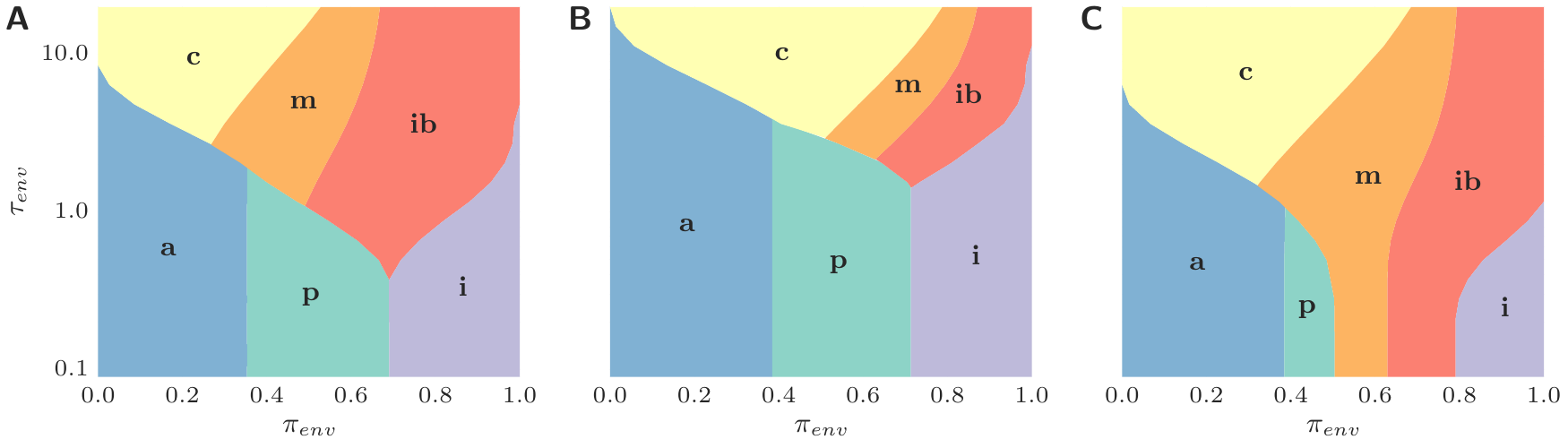}
    \caption{
    Phase diagrams for different costs as those assumed in Fig.~2.
    Phases represent \textbf{a}daptive, \textbf{p}roto-adaptive, \textbf{i}nnate, \textbf{i}nnate \textbf{b}et hedging, and \textbf{C}RISPR-like strategies as defined previously.
    The phase boundaries shift but the ordering of the phases is robust. 
        (A) More costly CRISPR and adapted defense, i.e. $c_{\rm constitutive} = \left(1.9 - c_{\rm defense}\right) / \left(c_{\rm defense} - 0.1\right); \; c(p_{\rm uptake}) = 0.2\times p_{\rm uptake} + 2\times p_{\rm uptake}^2$.
        (B) More permissive costs, i.e. all costs scaled down by a factor of 2.
        (C) Less permissive costs, i.e. all costs scaled up by a factor of 1.5.
    \label{figSIaltphases}}
\end{figure*}

\begin{figure*}
    \centering
    \includegraphics[width=\linewidth]{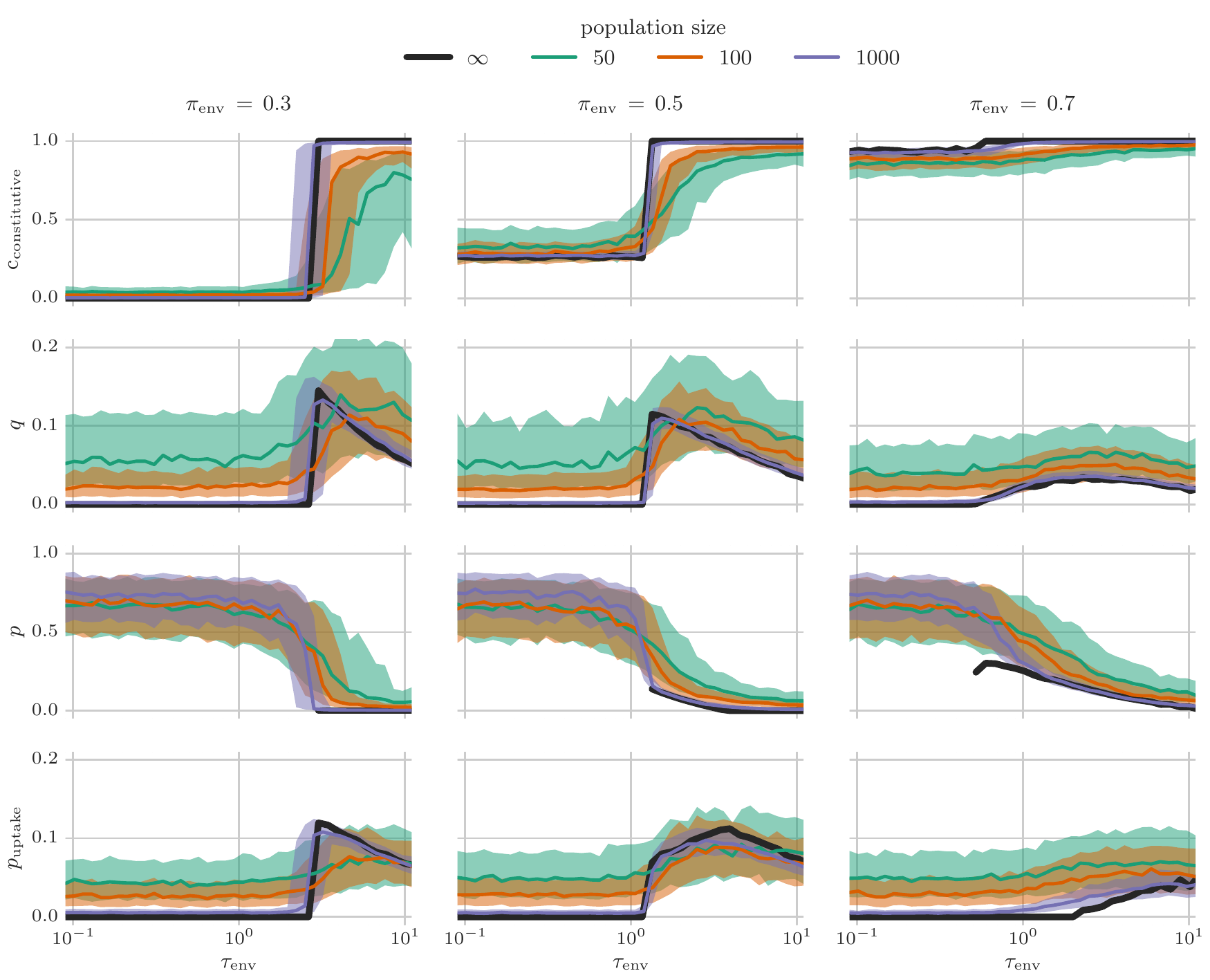}
    \caption{
        Influence of finite population size on optimal immune strategies from an agent-based simulation with evolving strategy parameters (switching rates and degree of adaptability) as described in Sec.~\ref{secfinitesize}.
        Subplots shows the median (solid line) and interquartile range (shaded area) of the strategy parameters at the end of a simulation of $100000$ generations length.
        Both are calculated from 500 independent simulations.
        In each simulation the strategy parameters evolve from a random initial distribution via mutation and selection.
        Mutations take place with a rate $0.01 \exp(-t/10000)$ per generation and are normally distributed with mean zero and standard deviation $0.25 \exp(-t/10000)$.
        The bound constraints on the parameters were enforced by setting the strategy parameters to the boundary value if outside after a mutation.
        Costs of different immune states as in Fig.~2.
    \label{figSIevol}}
\end{figure*}

\end{document}